\documentclass[12pt]{article}
\usepackage{epsfig,cite}
\usepackage {amsmath}
\usepackage{amssymb}
\include{epsf}
\newcount\timecount
\newcount\hours \newcount\minutes  \newcount\temp \newcount\pmhours
\hours = \time
\divide\hours by 60
\temp = \hours
\multiply\temp by 60
\minutes = \time
\advance\minutes by -\temp
\def\hour{\the\hours}
\def\minute{\ifnum\minutes<10 0\the\minutes
            \else\the\minutes\fi}
\def\clock{
\ifnum\hours=0 12:\minute\ AM
\else\ifnum\hours<12 \hour:\minute\ AM
      \else\ifnum\hours=12 12:\minute\ PM
            \else\ifnum\hours>12
                 \pmhours=\hours
                 \advance\pmhours by -12
                 \the\pmhours:\minute\ PM
                 \fi
            \fi
      \fi
\fi
}

\def\monthname{\relax\ifcase\month 0/\or January\or February\or
   March\or April\or May\or June\or July\or August\or September\or
   October\or November\or December\else\number\month/\fi}

\def\bold#1{\setbox0=\hbox{$#1$}%
     \kern-.025em\copy0\kern-\wd0
     \kern.05em\copy0\kern-\wd0
     \kern-.025em\raise.0433em\box0 }

\def\beq{\begin{equation}}
\def\eeq{\end{equation}}

\def\ga{\mathrel{\raise.3ex\hbox{$>$\kern-.75em\lower1ex\hbox{$\sim$}}}}
\def\la{\mathrel{\raise.3ex\hbox{$<$\kern-.75em\lower1ex\hbox{$\sim$}}}}
\def\gev{{\rm \, Ge\kern-0.125em V}}
\def\tev{{\rm \, Te\kern-0.125em V}}
\def\gyr{{\rm \, G\kern-0.125em yr}}

\def\tb{\tan \beta}

\def\gappeq{\mathrel{\rlap {\raise.5ex\hbox{$>$}}
{\lower.5ex\hbox{$\sim$}}}}
\def\lappeq{\mathrel{\rlap{\raise.5ex\hbox{$<$}}
{\lower.5ex\hbox{$\sim$}}}}
\def\Toprel#1\over#2{\mathrel{\mathop{#2}\limits^{#1}}}

\def\m12{m_{1\!/2}}

\def\bea{\begin{eqnarray}}
\def\eea{\end{eqnarray}}

\begin{document}
\begin{titlepage}
\pagestyle{empty}
\baselineskip=21pt
\rightline{KCL-PH-TH/2014-13, LCTS/2014-13, CERN-PH-TH/2014-060}
\vskip 1in
\begin{center}
{\large{\bf Closing in on the Tip of the CMSSM Stau Coannihilation Strip}}
\end{center}
\begin{center}
\vskip 0.5in
{\bf Nishita~Desai}~$^{1}$,
{\bf John~Ellis}~$^{2,3}$,
{\bf Feng Luo}~$^{2}$ and
{\bf Jad~Marrouche}~$^4$\\
\vskip 0.5in
{\small {\it
$^1${Institut f\"ur theoretische Physik, Universit\"at Heidelberg, \\
    Heidelberg 69120, Germany}\\
$^2${Theoretical Particle Physics and Cosmology Group, Department of
  Physics, \\ King's College London, London WC2R 2LS, United Kingdom}\\
$^3${Theory Division, CERN, CH-1211 Geneva 23, Switzerland}\\
$^4${Physics Department, CERN, CH-1211 Geneva 23, Switzerland}\\
}}

\vskip 0.75in
{\bf Abstract}
\end{center}
\baselineskip=18pt \noindent

Near the tip of the ${\tilde \tau}$ coannihilation strip in the CMSSM
with a neutralino LSP $\chi$, the astrophysical cold dark matter
density constraint forces the ${\tilde \tau} - \chi$ mass difference
to be small. If this mass difference is smaller than $m_\tau$, the ${\tilde
  \tau}$ may decay either in the outer part of an LHC detector - the
`disappearing track' signature - or be sufficiently long-lived to
leave the detector before decaying - the long-lived massive
charged-particle signature. We combine searches for these signatures
with conventional $E_T^{miss}$ searches during LHC Run~1, identifying
the small remaining parts of the CMSSM ${\tilde \tau}$ coannihilation
strip region that have not yet been excluded, and discussing how they
may be explored during Run~2 of the LHC.

\vfill
\end{titlepage}
\baselineskip=18pt

\section{Introduction}

A major theme of the searches for new physics during Run~1 of the LHC
at 7 and 8~TeV in the centre of mass has been the search for supersymmetry
via various experimental signatures~\cite{ATLASsusy,CMSsusy}. The constraints imposed by the
absences of any statistically-significant excesses of events with such signatures
are frequently interpreted assuming that $R$ parity is conserved, in which case
the lightest supersymmetric particle (LSP) could be present today as a cosmological
relic~\cite{ehnos} from the Big Bang, and many astrophysical constraints also come into play.
Foremost among these is the density of cold dark matter~\cite{wmap,planck}, which imposes
restrictions on supersymmetric model parameters that are often respected only
in narrow strips in the supersymmetric parameter space. This is the case, for
example, if it is assumed that the LSP is the lightest neutralino $\chi$. In
particular, in the minimal supersymmetric extension of the Standard Model
with soft supersymmetry-breaking parameters $m_{1/2}, m_0$ and $A_0$ assumed to be universal at the 
grand unification scale (the CMSSM)~\cite{funnel,cmssm}, the cold dark matter density constraint
is respected along strips where $\chi$ coannihilation with the lighter ${\tilde \tau}$
slepton~\cite{efo} or the lighter $\tilde t$ squark~\cite{stopchico} is important, or where $\chi - \chi$ annihilation
is enhanced by a direct-channel Higgs resonance~\cite{funnel,efgosi}, or an enhanced Higgsino
component in the composition of the LSP $\chi$~\cite{fp}.

As discussed in previous papers~\cite{celmov,dragon,stausoon}, the ${\tilde \tau} - \chi$ coannihilation strip
region in the CMSSM is particularly accessible to supersymmetry searches
at the LHC. However, complete exploration of this region requires a combination
of different search strategies. As the tip of the ${\tilde \tau} - \chi$ coannihilation strip
is approached at large $m_{1/2}$, the ${\tilde \tau} - \chi$ mass difference $\Delta m$
becomes very small. If $\Delta m > m_\tau$, the $\tilde \tau$ and other heavier
sparticles decay rapidly into the LSP $\chi$, providing a classical $E_T^{miss}$
signature. However, if $\Delta m < m_\tau$ the $\tilde \tau$ lifetime becomes so
long that it may decay in the outer part of a generic LHC detector - the
`disappearing track' signature - or even outside the detector altogether, in
which case the $\tilde \tau$ would appear as a slow-moving massive, penetrating
charged particle. Full exploration of the CMSSM ${\tilde \tau} - \chi$ 
coannihilation strip therefore requires a careful combination of searches for
these signatures as well as for $E_T^{miss}$.

In a previous paper~\cite{celmov}, the principal decay rates and the lifetime of the $\tilde \tau$ 
in the CMSSM when $\Delta m < m_\tau$ were re-evaluated, and the impact on the
CMSSM of the Run~1 LHC searches for massive metastable charged particles were
analyzed. Subsequently, updated LHC results from searches for such particles have
been made available~\cite{CMSex}, as well as searches for $E_T^{miss}$ events~\cite{ATLAS20} 
and disappearing tracks~\cite{ATLAS_disappearing}.
The purpose of this paper is to make a combined analysis of these different searches
within the CMSSM, identify the remaining regions of the CMSSM $\tilde \tau$
coannihilation strip, and discuss how they may be explored in Run~2 of the LHC.

In Section~2 of this paper we first review relevant features of the $\tilde \tau$
coannihilation strip region within the CMSSM, which extends up to $m_{1/2} \sim 1300$~GeV
for $\tan \beta = 40$ and $A_0 > 0$. We then review the calculations
of $\tilde \tau$ decays when $\Delta m < m_\tau$, which indicate that the
dominant $\tilde \tau$ signature would be a massive metastable charged particle if
$\Delta m \lappeq 1.2$~GeV and a disappearing track if $\Delta m \gappeq 1.2$~GeV.

In Section~3 we discuss the impacts of the relevant LHC Run~1 searches for new physics,
including regions where the relic LSP density is less than the total cold dark matter
density, as would be allowed if there is another component of the astrophysical cold
dark matter. We first discuss the $E_T^{miss}$ searches, 
which exclude the relevant portions of the CMSSM parameter space
where $\Delta m > m_\tau$ and $m_{1/2} < 780$~GeV. For
$\tan \beta = 10$, these searches exclude the portion of the $\tilde \tau$ coannihilation strip
where $\Delta m \gappeq 3$~GeV, whereas $\Delta m$ as large as 9~GeV can be allowed for
$\tan \beta = 40$. We then update our previous analysis of the metastable
$\tilde \tau$ case, finding that the most recent LHC Run~1 search for such particles
excludes $m_{1/2} \lappeq 800$~GeV to $\lappeq 1100$~GeV
for $\Delta m \lappeq 1.2$~GeV, depending on the value of $\tan \beta$ and $A_0$.
We then analyze the impact of the disappearing track search on the intermediate band where 
$m_\tau > \Delta m \gappeq 1.2$~GeV, using {\sc Pythia~8}~\cite{pythia8, py8susy} to
simulate $\tilde \tau$ decays in an LHC detector outside the beam-pipe.
We find that this search is weaker than the other constraints, yielding $m_{1/2} \gappeq 400$~GeV.

In Section~4 we discuss the interplay of these different searches, as well as the 
constraints from the observed value of the Higgs mass $m_h$~\cite{ATLASH,CMSH}, calculated using
the recently-released {\tt FeynHiggs~2.10.0}~\cite{feynhiggs2.10.0}.  

In Section~5 we consider the sensitivities of LHC Run~2 searches with 300/fb of
integrated luminosity at 14~TeV in the centre of mass. The conventional $E_T^{miss}$
searches should have sufficient sensitivity to find evidence for supersymmetry or
to exclude the coannihilation region of the CMSSM if $\Delta m > m_\tau$.
Likewise, searches for massive metastable charged particles should be able to
find evidence for the $\tilde \tau$ or to exclude the coannihilation region of the 
CMSSM if $\Delta m \lappeq 1.2$~GeV. However, simple extrapolation of the
current disappearing track searches indicates that they would have insufficient
sensitivity to exclude or find evidence for supersymmetry if $m_\tau > \Delta m \gappeq 1.2$~GeV,
so we consider ways in which the sensitivity of future such searches could be enhanced.

Finally, Section~6 summarizes our conclusions.

\section{The $\tilde \tau$ Coannihilation Strip and its Decays within the CMSSM}

\subsection{Anatomy of the Stau Coannihilation Strip Region}
\, 
The focus of our attention in this paper is the CMSSM, in which $R$ parity is
conserved and it is assumed that universal soft
supersymmetry-breaking parameters $m_{1/2}, m_0$ and $A_0$ are input at the GUT scale.
We assume that the stable LSP is the lightest neutralino $\chi$, giving priority to the 
CMSSM parameter region near the strip where
its astrophysical relic density is brought into the range $0.115 < \Omega_\chi h^2 < 0.125$~\cite{planck}
that is acceptable within conventional
cosmology by coannihilation with the lighter tau-slepton $\tilde \tau$ and other, heavier
sleptons, but also considering smaller values of $\Delta m$ that yield lower values of
$\Omega_\chi h^2$. Our objective is to study the extent to which this simplest supersymmetric
scenario has been explored with data from Run~1 of the LHC at 7 and 8~TeV in the centre of
mass, and the extent to which it can be explored further with future LHC data at 14~TeV.
As we discuss, even this simplest scenario has rich phenomenological possibilities
beyond the standard $E_T^{miss}$ signatures, posing challenges for its complete exploration.

\begin{figure}[h!]
\centering
\epsfig{file=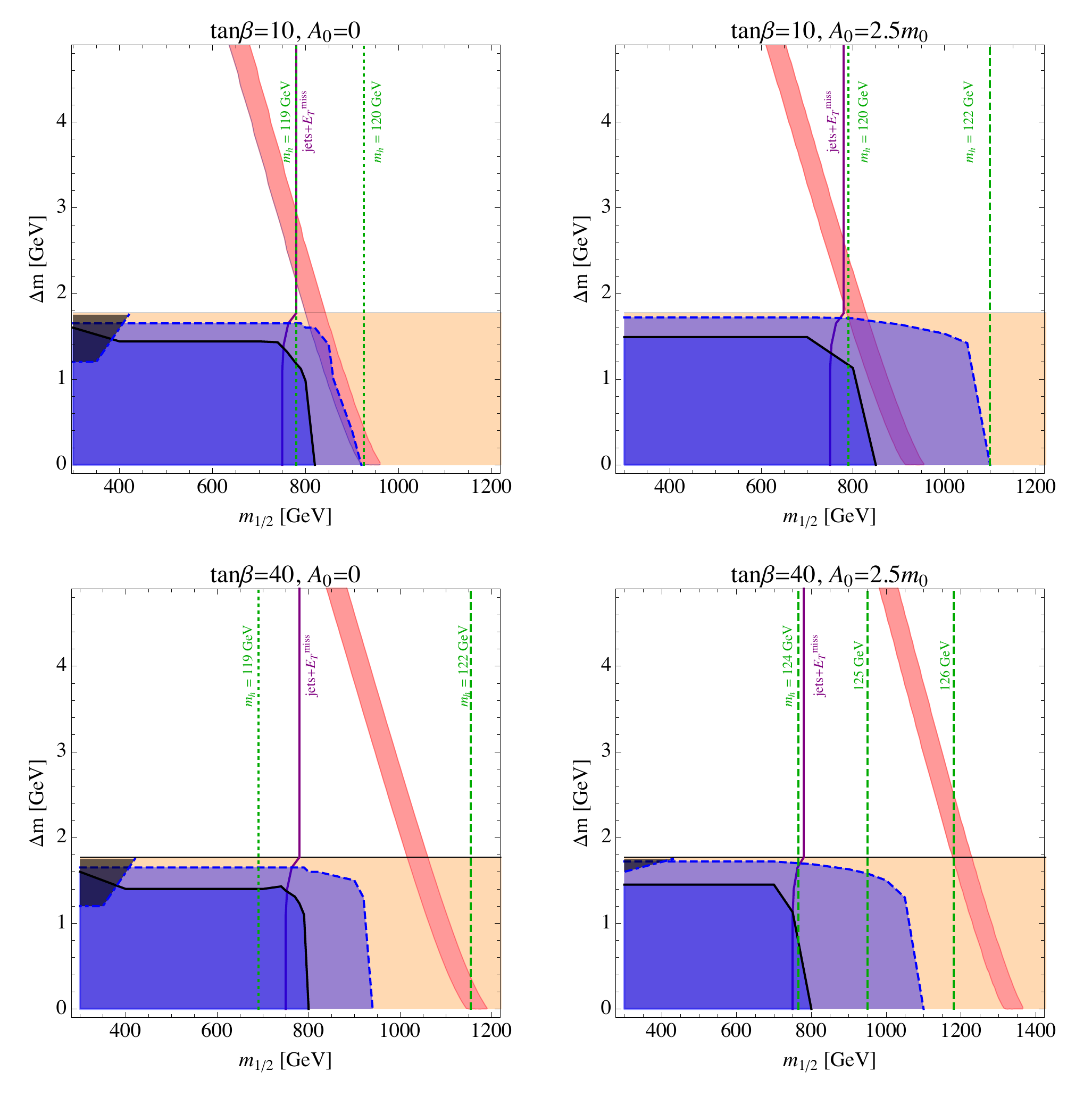,height=6in}
\caption{
{\it Overviews of the regions of the CMSSM parameter space
with small mass difference $\Delta m \equiv m_{\tilde \tau}-m_\chi$ 
for $\tb = 10$ (upper panels) and $\tb = 40$ (lower panels), 
and for $A_0 = 0$ (left panels) and $2.5 \, m_0$ (right panels).
The bands with $\Delta m < m_\tau$ are shaded beige,
and the coannihilation strips where $0.125 > \Omega_\chi h^2 > 0.115$
as calculated using {\tt SoftSUSY~3.3.7}~\protect\cite{softsusy} coupled to 
{\tt MicrOMEGAs 3.5.5}~\protect\cite{micromegas_mssm} are shaded pink.
The lower limit on $m_{1/2}$ from the ATLAS $E_T^{miss}$ search during Run~1 at the LHC~\protect\cite{ATLAS20}
is represented in each panel by a maroon line, and contours of $m_h$ calculated using
{\tt FeynHiggs~2.10.0}~\protect\cite{feynhiggs2.10.0} are shown as green (dashed or dotted) lines.
Parameter regions excluded by searches for the direct and total
production of metastable charged particles~\protect\cite{CMSex} are shaded darker and
lighter blue, respectively, and regions excluded by searches for particles leaving disappearing tracks are shaded
grey (see Section~3 for details).
}} 
\label{fig:DeltaM}
\end{figure}

As is well-known, as $m_{1/2}$ increases toward the tip of the stau coannihilation strip the 
${\tilde \tau} - \chi$ mass difference $\Delta m$ decreases monotonically towards zero, which is attained at
$m_{1/2} = {\cal O}(1000)$~GeV, the maximum value of $m_{1/2}$ depending
on the values of $\tan \beta$ and $A_0$. In this paper, we use consistently 
{\tt SoftSUSY~3.3.7}~\cite{softsusy} 
to calculate the sparticle spectrum, and the latter is passed to {\tt MicrOMEGAs 3.5.5}~\cite{micromegas_mssm} to calculate
$\Omega_\chi h^2$. Fig.~\ref{fig:DeltaM} displays
bands with $\Delta m \le 5$~GeV for values of $m_{1/2}$ close to the tips of
the coannihilation strips for $\tan \beta = 10$ (upper panels) and 40 (lower panels), in each case
for the two choices $A_0 = 0$ (left panels) and $2.5 \, m_0$ (right panels). The choices of $\tan \beta$ are
representative of the larger and smaller values found in the coannihilation region in a
recent global analysis of the CMSSM parameter space~\cite{MC8}, and the restriction to $A_0 \ge 0$
is motivated by the Higgs boson mass $m_h$ measured at the LHC, which is easier to reproduce
for positive values of $A_0$~\footnote{These choices of $\tan \beta$ and $A_0$ are also used in~\cite{celmov}.
However, to avoid confusion when the reader compares 
the results of this paper with the previous one~\cite{celmov}, where the {\tt SSARD} code~\cite{SSARD} was used,
please note that here we use the opposite convention for the sign of $A_0$.}. 
The coannihilation strips where $0.115 < \Omega_\chi h^2 < 0.125$
are shown as pink bands. We see that the
strips for $\tan \beta = 10$ terminate when $\Delta m \to 0$ at $m_{1/2} \simeq 900$ to 950~GeV,
with little sensitivity to $A_0$, whereas the strips for $\tan \beta = 40$ and $A_0 = 0 \, \, (2.5 \, m_0)$
extend to larger $m_{1/2} \simeq 1150$ to 1200~GeV (1300 to 1350~GeV). We also see that $\Delta m$ drops
below $m_\tau$ for $m_{1/2} \simeq 800$ to 850~GeV for $\tan \beta = 10$,
and $m_{1/2} \simeq 1050$ to 1100~GeV (1200 to 1250~GeV) for $\tan \beta = 40$, respectively.

The strips within which the relic LSP density $\Omega_\chi h^2$ falls inside the range allowed by the
available astrophysical and cosmological data for the total cold dark matter density
$\Omega_{CDM} h^2$ are quite narrow, since $\Omega_{CDM} h^2$ is tightly constrained,
at the \% level, and the theoretical uncertainties in calculating $\Omega_\chi h^2$ are
small within conventional Big Bang cosmology. However, we note that $\Omega_\chi h^2$
could be substantially smaller if the LSP is not the only important component of the cold dark matter.
We therefore consider also the regions of the CMSSM parameter space with lower 
$\Omega_\chi h^2$ that have $\Delta m$ smaller than along the coannihilation strips
displayed in Fig.~\ref{fig:DeltaM}.

We return later to the other features exhibited in Fig.~\ref{fig:DeltaM}.

\subsection{Review of $\tilde \tau$ Decays}

The starting point for the calculation of $\tilde \tau$ decays in the CMSSM
is the ${\tilde \tau} - \chi - \tau$ Lagrangian, which was derived in detail in the 
Appendix of~\cite{celmov}. If $\Delta m > m_\tau$, the dominant decay mode is 
${\tilde \tau} ^-\to \tau^- \chi$, which gives a $\tilde \tau$ lifetime many orders of magnitude smaller 
than 1 nanosecond. If $\Delta m < m_\tau$, the three- and four-body decay modes, 
${\tilde \tau}^- \to a_1^-(1260) \nu_\tau \chi$, 
${\tilde \tau}^- \to \rho^-(770) \nu_\tau \chi$, 
${\tilde \tau}^- \to \pi^- \nu_\tau \chi$,
${\tilde \tau}^- \to \mu^- {\bar \nu_\mu} \nu_\tau \chi$ and 
${\tilde \tau}^- \to e^- {\bar \nu_e} \nu_\tau \chi$, 
are the relevant ones, and their branching ratios varying with $\Delta m$, in particular. 
These channels close in sequence toward the tip of the coannihilation strip as $\Delta m \to 0$.
Analytic expressions of the $\tilde \tau$ decay rates for these channels can be found in the 
Appendix of~\cite{celmov}.

For $1.2~\text{GeV} \lappeq \Delta m < m_\tau$, the $\tilde \tau$ lifetime is between order 
one and several hundred nanoseconds, so that the $\tilde \tau$ may decay in the outer part of 
the ATLAS and CMS detectors, and a disappearing track is the dominant $\tilde \tau$ signature. Over most 
of this range of $\Delta m$, ${\tilde \tau}^- \to \rho^-(770) \nu_\tau \chi$ is the dominant decay mode with 
a branching ratio varying between $\sim 29\%$ and $\sim 37\%$. The ${\tilde \tau}^- \to \pi^- \nu_\tau \chi$ branching ratio
increases roughly linearly from $\sim 13\%$ to $\sim 36\%$ with the decrease of $\Delta m$, and it becomes the
dominant mode at the lower end of $\Delta m$. The branching ratio of 
${\tilde \tau}^- \to e^- {\bar \nu_e} \nu_\tau \chi$ is $\sim 18\%$ to $20\%$ over this $\Delta m$ range, and 
${\tilde \tau}^- \to \mu^- {\bar \nu_\mu} \nu_\tau \chi$ is $\sim 1\%$ smaller than the former. 
The ${\tilde \tau}^- \to a_1^-(1260) \nu_\tau \chi$ branching ratio is about the same size as each of the four-body 
decay modes at $\Delta m \sim m_\tau$, decreasing to $\sim 5\%$ at $\Delta m \sim 1.5$ GeV, and 
continually decreasing until its phase space vanishes at $\Delta m = m_{a_1}$. 
For $\Delta m \lappeq 1.2$ GeV, the $\tilde \tau$ is sufficiently long-lived so that it leaves the detectors before 
decaying, and the signature would be a massive metastable charged particle. 

\begin{figure}[h!]
\vskip 0.5in
\hspace*{-.30in}
\begin{minipage}{8in}
\epsfig{file=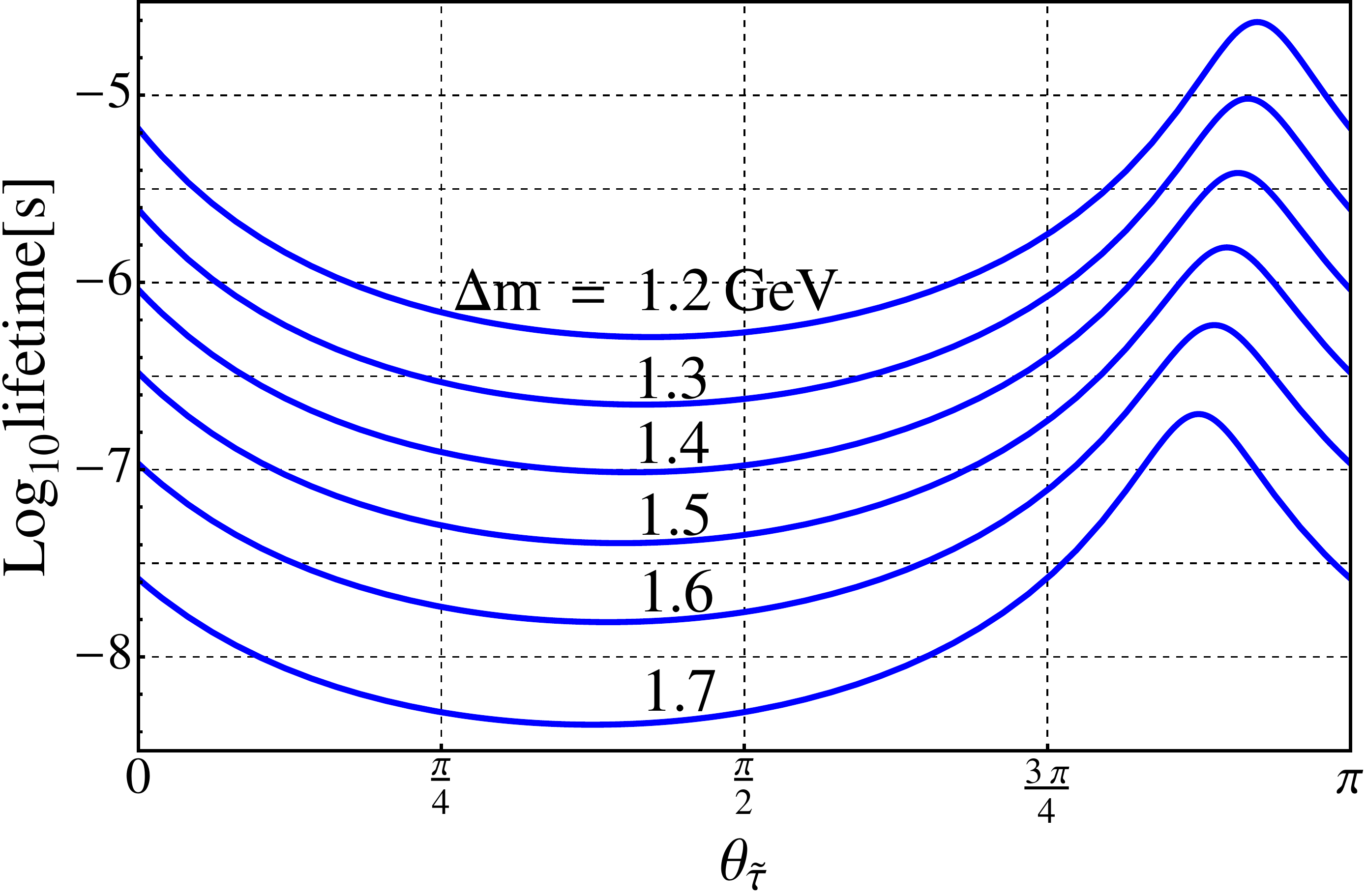,height=2.3in}
\epsfig{file=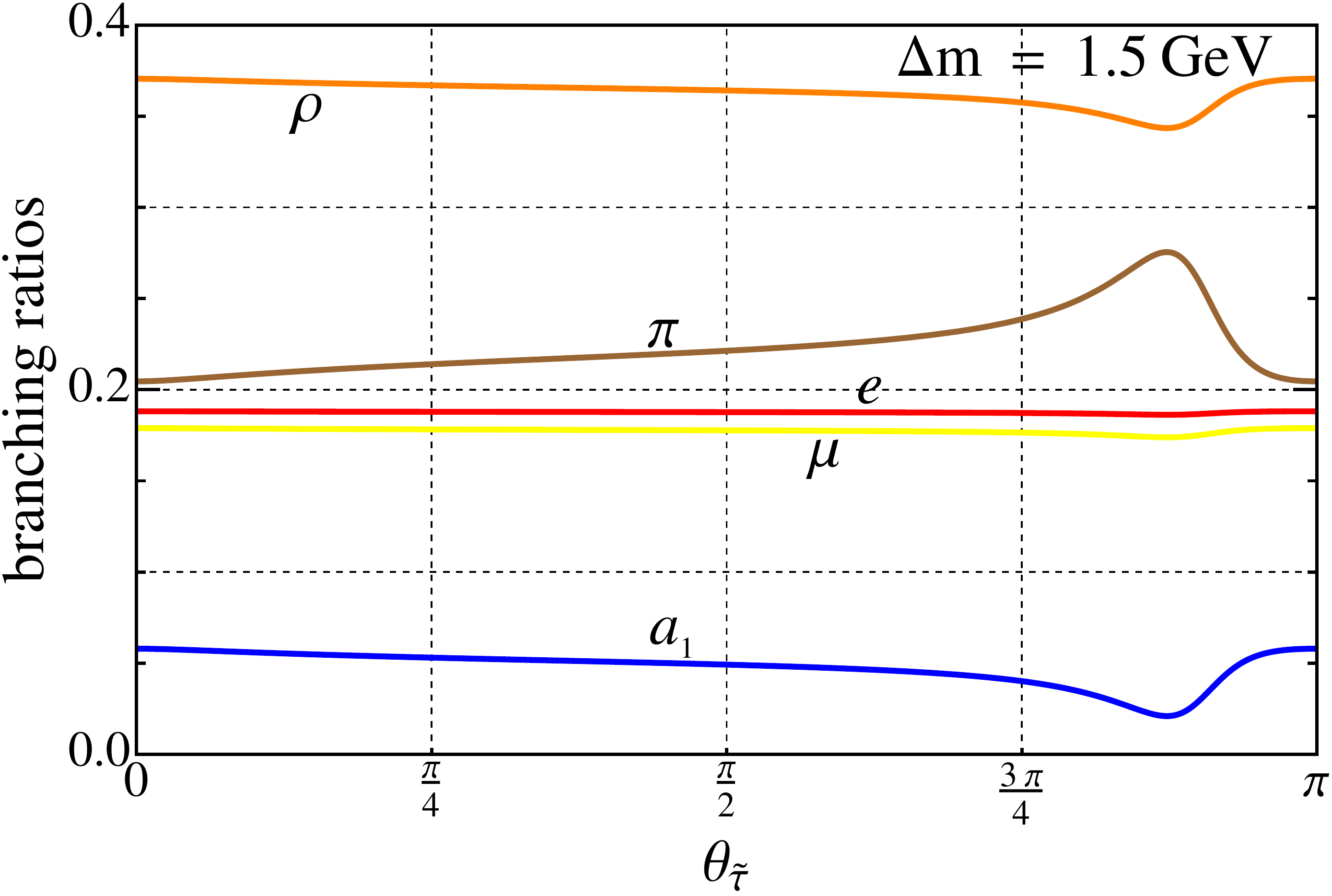,height=2.3in}
\end{minipage}
\caption{
{\it Left panel: the $\tilde \tau$ lifetime as a function of the ${\tilde \tau_L} - {\tilde \tau_R}$ mixing angle, 
$\theta_{\tilde \tau}$, for $\Delta m = 1.2, 1.3, \cdots, 1.7$ GeV.  
Right panel: the $\tilde \tau$ decay branching ratios as functions of $\theta_{\tilde \tau}$ 
for $\Delta m = 1.5$ GeV. The blue, orange, brown, yellow, and red lines 
are for the final states with $a_1(1260)$, $\rho(770)$, $\pi$, $\mu$, and $e$, respectively,
indicated by the labels adjacent to the corresponding curves. 
In both panels, a pure bino-like $\chi$ with a mass of 300 GeV is assumed. 
}} 
\label{fig: stau_mixing_angle_dependence}
\end{figure}

The $\tilde \tau$ lifetime and decay branching ratios were plotted in Fig.~7 and 8 in~\cite{celmov}. 
In those figures the $\chi$ was assumed to be pure bino-like with a mass of 300 GeV, 
and the ${\tilde \tau_L} - {\tilde \tau_R}$ mixing angle, $\theta_{\tilde \tau}$, was taken to be $\pi/3$. These parameters were 
chosen in order to compare with the results of an earlier paper~\cite{Jittoh} where some differences in the $\tilde \tau$ decay 
calculations were found. We note that along $\tilde \tau$ coannihilation strip within the CMSSM, the $\chi$ is almost 
a bino, and the $\tilde \tau$ is almost right-handed. To see the effects of $\theta_{\tilde \tau}$ on the 
$\tilde \tau$ lifetime and decay branching ratios, we plot in the left panel of Fig.~\ref{fig: stau_mixing_angle_dependence} 
the $\tilde \tau$ lifetime as a function of $\theta_{\tilde \tau}$ for the same 300 GeV pure bino-like $\chi$ and for 
$\Delta m = 1.2, 1.3, \cdots, 1.7$ GeV, and in the right panel we show the 
branching ratios $vs.$ $\theta_{\tilde \tau}$ curves for the same $\chi$ parameters and for 
$\Delta m = 1.5$ GeV. In these plots, $\theta_{\tilde \tau} = \pi /2$ corresponds to a pure right-handed $\tilde \tau$.
We only show the plots for $\theta_{\tilde \tau} \in [0, \pi]$ because adding a $\pi$ to $\theta_{\tilde \tau}$ is equivalent to 
change the overall sign of the ${\tilde \tau} - \chi - \tau$ Lagrangian, and it has no effect
for the $\tilde \tau$ decay rates calculations performed in~\cite{celmov}. 
We can see that the $\tilde \tau$ lifetime strongly depends on $\theta_{\tilde \tau}$, 
but this dependence is not as strong as that of on $\Delta m$. On the other hand, the branching ratios only mildly
depend on $\theta_{\tilde \tau}$, and this is also true for other choices of $\Delta m$ which we do not show here. 
Finally, we note that for a given $\Delta m$, the $\tilde \tau$ lifetime is roughly proportional to $m_\chi$ (so that 
it is roughly proportional to $m_{1/2}$ in the CMSSM due to the relation $m_\chi \sim 0.42 \, m_{1/2}$), while the 
branching ratios are not sensitive to $m_\chi$. 

\begin{figure}[h!]
\vskip 0.5in
\hspace*{1.0in}
\begin{minipage}{8in}
\epsfig{file=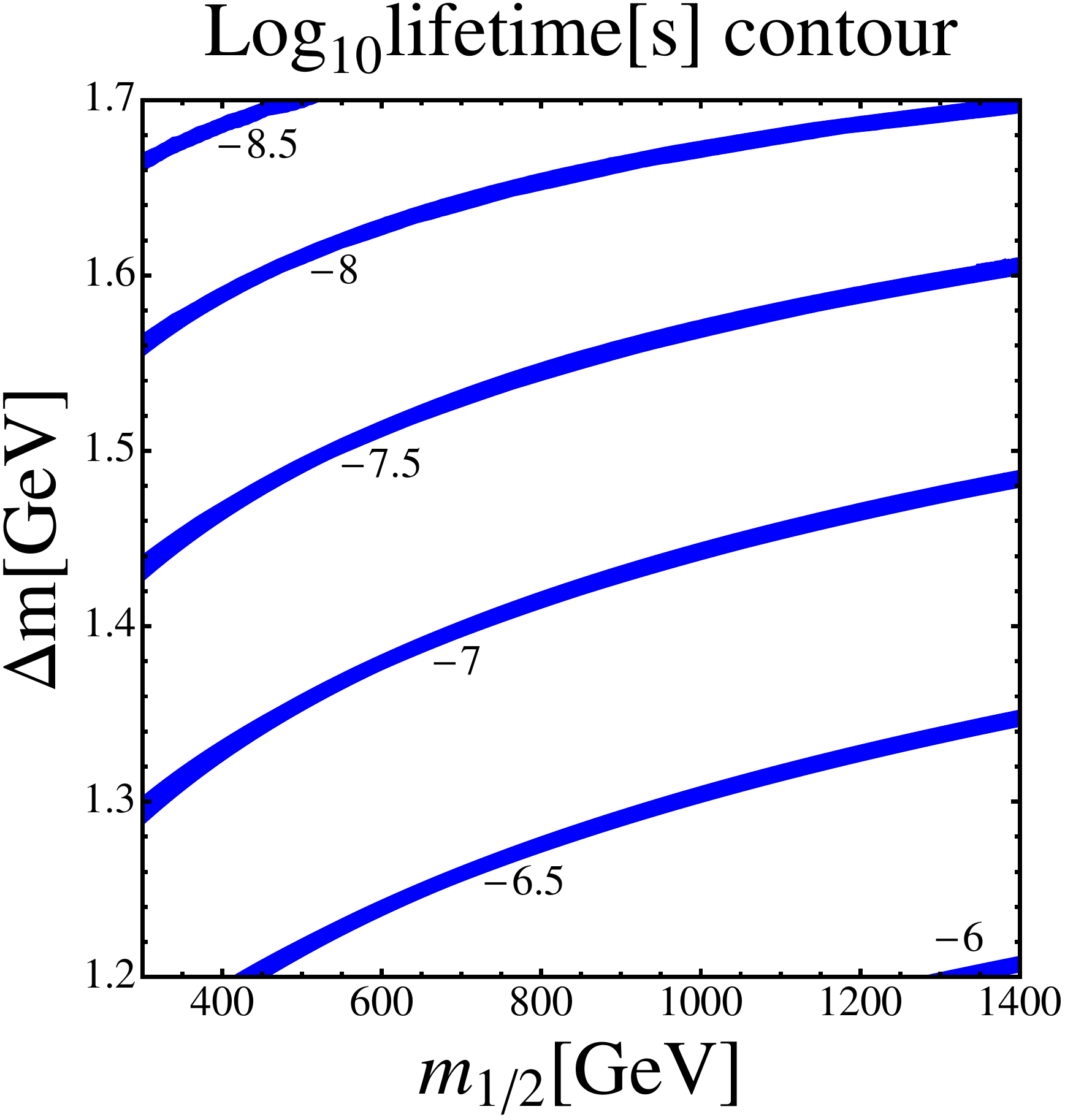,height=4.3in}
\end{minipage}
\caption{
{\it Some $\tilde \tau$ lifetime contours as functions of $m_{1/2}$ and
$\Delta m$, for the four choices of the CMSSM parameters used in Fig.~\protect\ref{fig:DeltaM}, namely, 
$\tb = 10, 40$ and $A_0 = 0, 2.5 \, m_0$. The widths of the contours span ranges of the $\tilde \tau$ lifetime 
for the different choices of $\tb$ and $A_0$. 
}}
\label{fig: stau_lifetime_contour}
\end{figure}

Fig.~\ref{fig: stau_lifetime_contour} shows the $\tilde \tau$ lifetime contours as functions of $m_{1/2}$ and
$\Delta m$, for the range $m_{1/2} \in (300, 1400)$ GeV and $\Delta m \in (1.2, 1.7)$ GeV. For all the four choices 
of the CMSSM parameters used in Fig.~\ref{fig:DeltaM}, within this small $\Delta m$ 
range the $\chi$ is almost a bino, and the $\tilde \tau$ is almost right-handed.  Therefore, the $\tilde \tau$ lifetime 
essentially only depends on $\Delta m$ and $m_\chi$ (or, equivalently, $m_{1/2}$), so that the contours are almost identical 
for all the four choices of the CMSSM parameters. The widths of the contours in
Fig.~\ref{fig: stau_lifetime_contour} span the dependence of the $\tilde \tau$ lifetime on 
$\tb$ and $A_0$. One can check that the $\tilde \tau$ lifetimes at $m_{1/2} \sim 720$ GeV are consistent with the 
values shown in the left panel of Fig.~\ref{fig: stau_mixing_angle_dependence} for $\theta_{\tilde \tau} \sim \pi /2$. 

\section{The Impacts of LHC Run~1 Searches on the Coannihilation Strip Region}

\subsection{Searches for $E_T^{miss}$ Events}

\begin{figure}[t]
\vskip 0.5in
\hspace*{-0.65in}
\begin{minipage}{8in}
\epsfig{file=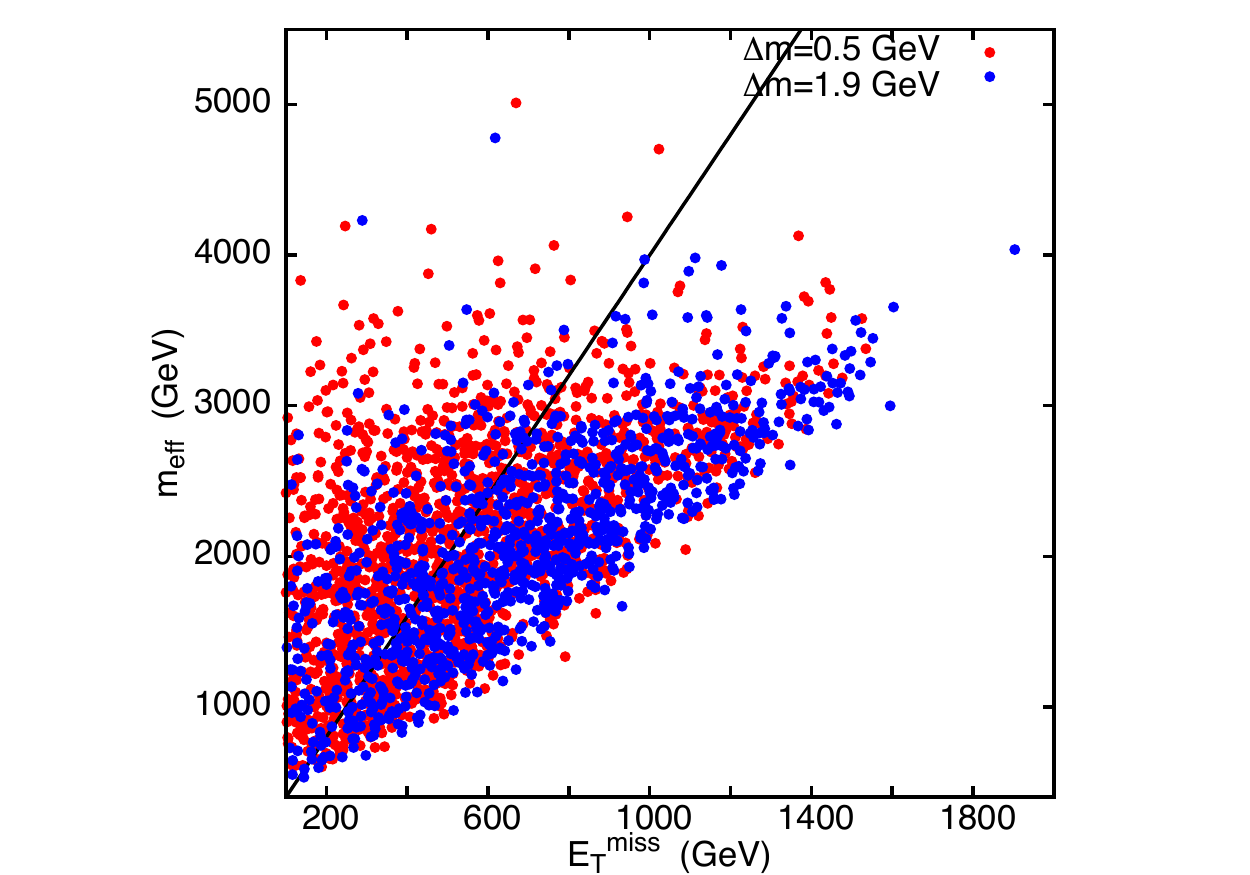,height=3in}
\hspace*{-1.0in}
\epsfig{file=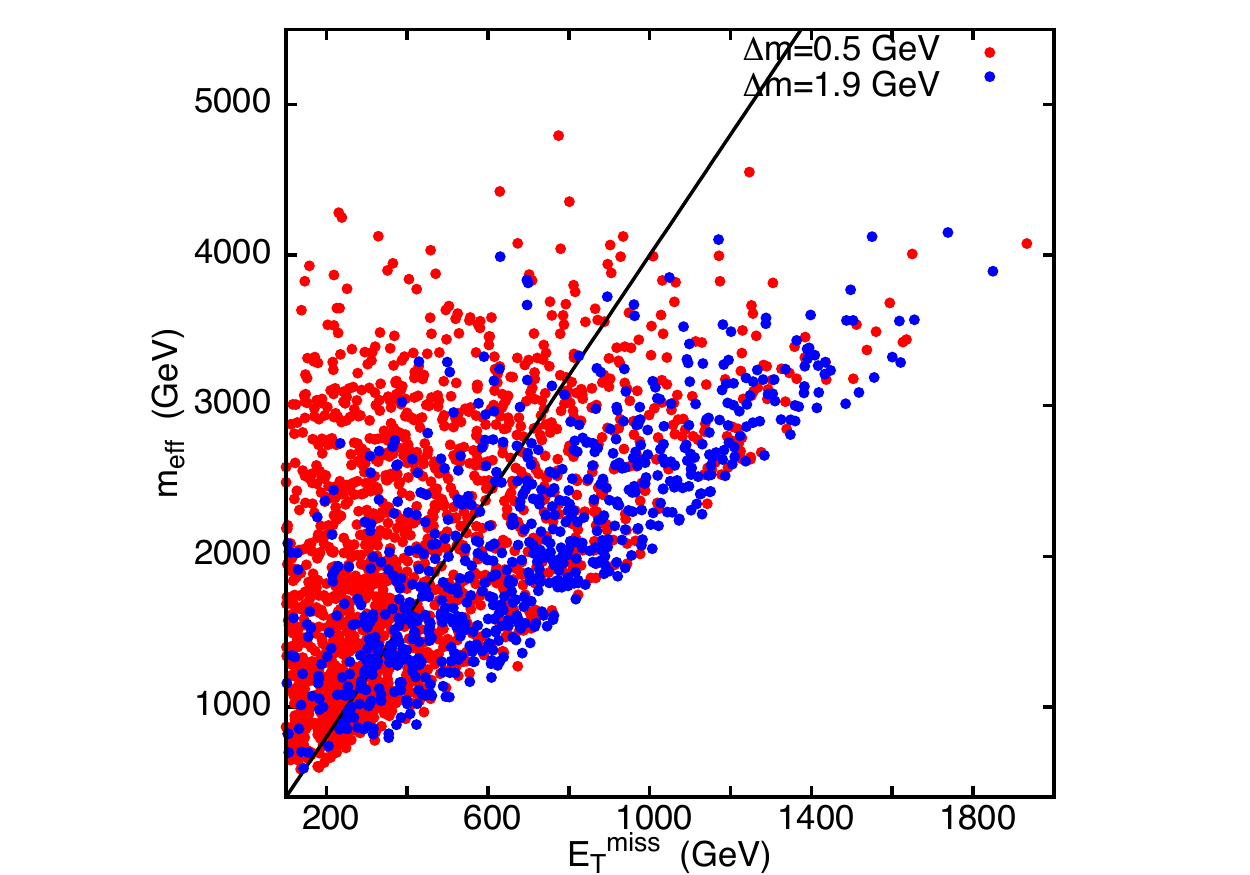,height=3in}
\end{minipage}
\caption{
{\it Scatter plot of $E_T^{miss}$ and $m_{eff}$ in the four-jet channel for CMSSM scenarios with metastable staus 
($\Delta m =0.5$ GeV, red points) and with rapid ${\tilde \tau} \to \tau + \chi$ decays ($\Delta m =1.9$ GeV, blue points).
The left plot is for $\tan \beta = 10$, $A_0 = 0$ and the right plot is for $\tan \beta = 40$, $A_0 = 2.5\,m_0$, both with
$m_{1/2} = 800$~GeV.  The solid diagonal lines correspond to the ATLAS cut 
$E_T^{\mathrm{miss}} > 0.25\, m_{\mathrm{eff}}$~\protect\cite{ATLAS20}.
}}
\label{fig:met-meff}
\end{figure}

Both the ATLAS and CMS collaborations have performed dedicated sets of experimental searches for 
generic new physics signatures with an abundance of missing transverse energy, $E_T^{miss}$,
as motivated in particular by supersymmetric
models in which the stable lightest supersymmetric particle, commonly chosen to be the neutralino,
is a massive dark matter particle. The signatures studied generally include jets,
which could originate, e.g., from the pair production and subsequent cascade decays of squarks and gluinos. 
These searches have been carried out for a range of different final states,
some including reconstructed leptons as well as jets tagged as originating from $b$-quarks, 
for a number of different ranges of the missing transverse energy and the total transverse energy. 
None of these searches found any significant evidence for new physics exhibiting these signatures in the LHC Run 1 data.

The ATLAS collaboration has provided an interpretation of their data in the context of the CMSSM based on the 
2012 dataset of 20/fb at a centre of mass energy of 8~TeV~\cite{ATLAS20}. 
The interpretation is presented in the $(m_0, m_{1/2})$ plane for a fixed value of $\tan \beta = 30$
and $A_0 = 2 \, m_0$ (in our convention for the sign of $A_0$). 
Several different searches have been discussed in~\cite{ATLAS20}, but for the purposes of our study we 
concentrate on the 0-lepton search with 2-6 jets, as this provides the most stringent limit in the 
region of the stau coannihilation strip, and is also relatively insensitive to the values of 
$\tan \beta$ and $A_0$, as shown in a previous study~\cite{MC8}. 
As is shown by ATLAS, the CMSSM interpretation of this search provides a limit
$m_{1/2} > 780$~GeV at the 95\% CL near the stau coannihilation strip where $\Delta m > m_\tau$~\footnote{We
note that the analogous CMS $E_T^{miss}$ analysis could provide a similar sensitivity to the CMSSM
parameters, but a CMSSM interpretation is not provided.}.

We reproduce the ATLAS analysis in \cite{ATLAS20} using {\sc Pythia 8} with realistic smearing functions~\cite{CMS-DP} to take into account detector effects,
paying particular attention to the signal efficiencies of simulated points in the CMSSM parameter space 
close to the stau-coannihilation strip. We have considered the various signal regions defined for the ATLAS search, 
and studied how the different signal region efficiencies change as functions of the stau-neutralino mass splitting.  

In the stau-coannihilation region, we find the strongest exclusions in the three-jet channel ($3j$) and four-jet channels ($4jt$), 
in agreement with the ATLAS analysis \cite{ATLAS20}. Our simulation results in a limit on $m_{1/2} > (780, 830)$~GeV in the signal regions $3j$ and $4jt$ (as defined in \cite{ATLAS20}) for the same choice of parameters and agrees well with the upper limits of $m_{1/2} > (780, 840)$~GeV reported in the conference note \cite{ATLASnote} together with the analysis referred above. Combining this study with knowledge of the cross-section, we
extrapolate the ATLAS limit into the region where $\Delta m < m_\tau$.

This search requires tight cuts on missing transverse energy ($E_T^{miss}$) 
and the effective mass ($m_{eff}$) defined as the sum of the $p_T$ of all jets plus the $E_T^{miss}$.  
To investigate the variation in sensitivity of this search channel, 
we compare the distributions in these two variables for $\Delta m$ above and below $m_\tau$.
As can be seen in the scatter plot in Fig.~\ref{fig:met-meff}, there are events satisfying the ATLAS $4jt$
cuts even when $\Delta m = 0.5$~GeV and the stau is nearly stable (red points),
albeit with a smaller efficiency than for $\Delta m =1.9$~GeV (blue points).  A similar effect is observed in the $3j$ signal region on which ATLAS exclusions in this region are based.  
We quantify in Section~\ref{sec:combo} the sensitivity of this search for small $\Delta m$, in conjunction with
the other searches discussed in Sections~\ref{sec:MCPs} and \ref{sec:diss-track}.

\subsection{Searches for Metastable Charged Particles}
\label{sec:MCPs}

For stau lifetimes longer than a few nanoseconds, a significant fraction of staus 
lives long enough to escape the detector 
leaving a charged track signature.  We can also use the 95\% upper limits on long-lived charged particles 
at 8~TeV reported by CMS \cite{CMSex} to determine exclusions for small stau-neutralino mass differences $\Delta m$.  
The upper limits for direct production of stau pairs are between 4.3 fb for a stau mass of 126 GeV and 0.26 fb at 494 GeV, 
after which the limit plateaus.  The limits for direct + indirect production (i.e., via cascade decays of other 
sparticles) are similar.  For masses between 126-308 GeV, we interpolate using the 
discrete set of values given in~\cite{CMSex},
and beyond this we assume an upper limit of 0.3 fb, as suggested by the middle left panel of Fig.~8 of~\cite{CMSex}.    

We calculate the fraction of events with at least one stau that is long-lived enough to escape the CMS detector,
but exits within the central region $|\eta| < 2.1$ so that a track would be visible.   Since the CMS upper limit 
constrains the total cross section, applying this limit to the restricted range of $|\eta|$ is slightly conservative.  
We find that a significant fraction of the staus with $\Delta m < 1.4$ GeV are stable on the scale of the CMS detector.  
We can rule out all $m_{1/2}$ values up to about 700 GeV, corresponding to $m_\chi = 291$ GeV,
for $\Delta m < 1.4$ GeV by the direct production constraint.  The excusion tapers off for $\Delta m > 1.4$~GeV,
as can be seen in Fig.~\ref{fig:DeltaM}. The maximum $m_{1/2}$ ruled out at low $\Delta m$ is between 800-850 GeV,  
which corresponds to a stau mass of 336-345 GeV.  This agrees with the lower limit of
339 GeV for the mass of a stable stau established by the CMS direct search constraint.  

When looking at both direct and indirect production, the search is sensitive to $\Delta m < 1.6$~GeV
for $A_0 = 0$ and $\Delta m < 1.7$ GeV for $A_0 = 2.5\, m_0$.  The maximum $m_{1/2}$ exclusion
is between $930$ GeV (for $\tan \beta = 10, A_0 = 0$) and $1100$ GeV for both values of $\tan \beta$ with $A_0 = 2.5 \, m_0$.  
This corresponds to stau masses between 385-447 GeV, and is conservative compared to the corresponding upper limit 
reported by CMS of 500~GeV.

\subsection{Searches for Disappearing Tracks}
\label{sec:diss-track}

The disappearing track search by the ATLAS collaboration looks for well-defined tracks that do not proceed 
beyond the transition radiation tracker (TRT) region of the detector.  This corresponds to  a radial range of 
563-1066 mm and a pseudo-rapidity range of $|\eta| < 2.0 $.  We simulate all event selection cuts in \cite{ATLAS_disappearing},
namely (1) $E_T^{miss} > 70$ GeV, (2) $p_T^{jet_1} > 80$ GeV and 
(3) $\Delta \phi_\mathrm{min}^{\mathrm{jet}-E_T^{miss}} > 1.0$.  Several further cuts are applied to the stau track --- 
(1) We require  the track to be isolated by demanding that the sum of the $p_T$ of all charged tracks 
within a cone of 0.4 around the stau track is less than 0.04 times the $p_T$ of the track; 
(2) The candidate track has $p_T > 15$ GeV and is the highest $p_T$ track in the event; 
(3) The track has pseudo-rapidity in the range $0.1<|\eta|<1.9$.  For the disappearing track criterion, 
we demand that the stau decays within the radial and pseudo-rapidity range of the TRT detector.

After applying all the cuts, we validate our simulation by reproducing to within 10\% the efficiency for the benchmark 
Anomaly Mediated Supersymmetry-Breaking (AMSB) point reported in the analysis.  As seen from
Fig.~\ref{fig: stau_lifetime_contour},  a value of $\Delta m$ between 1.4 and 1.77 GeV results in stau lifetimes between
1 and 100 ns, which is the ideal range for disappearing-track signatures at the LHC.   

The 95\% cross-section upper limits reported by ATLAS for events with $p_T^{\mathrm{track}} > 75, 100, 150$ and 
$200$ GeV are 1.76 fb, 1.02 fb, 0.62 fb and 0.44 fb respectively, which we apply to the cross section for events
passing all the cuts (1), (2) and (3) enumerated above.  We find that restricting to direct-stau production does not
yield any exclusions for $m_{1/2} > 300$ GeV.  However, including both direct and indirect production, 
a small region below $m_{1/2} =400$ GeV is ruled out for $\Delta m >1.2 $ GeV for $A_0 = 0$ and for  
$\Delta m >1.6$ GeV for $A_0 = 2.5 \, m_0$.  The jets + $E_T^{miss}$ search described above 
provides much stronger constraints for such values of $\Delta m$, excluding $m_{1/2} < 760$ GeV in this region.

\section{Combination of LHC Constraints}
\label{sec:combo}

We now discuss the interplay of the various LHC constraints displayed in Fig.~\ref{fig:DeltaM}.
The solid maroon lines mark the boundary of the region still allowed following
the ATLAS $E_T^{miss}$ searches at 8~TeV. As discussed previously, the most relevant
search is that for jets + $E_T^{miss}$, which provides a limit at $m_{1/2} = 780$~GeV in the
region where $\Delta m > m_\tau$. This constraint is weakened when $\Delta m < m_\tau$,
as discussed earlier.  

\begin{figure}[ht]
\vskip 0.5in
\hspace*{-0.1in}
\begin{minipage}{8in}
\epsfig{file=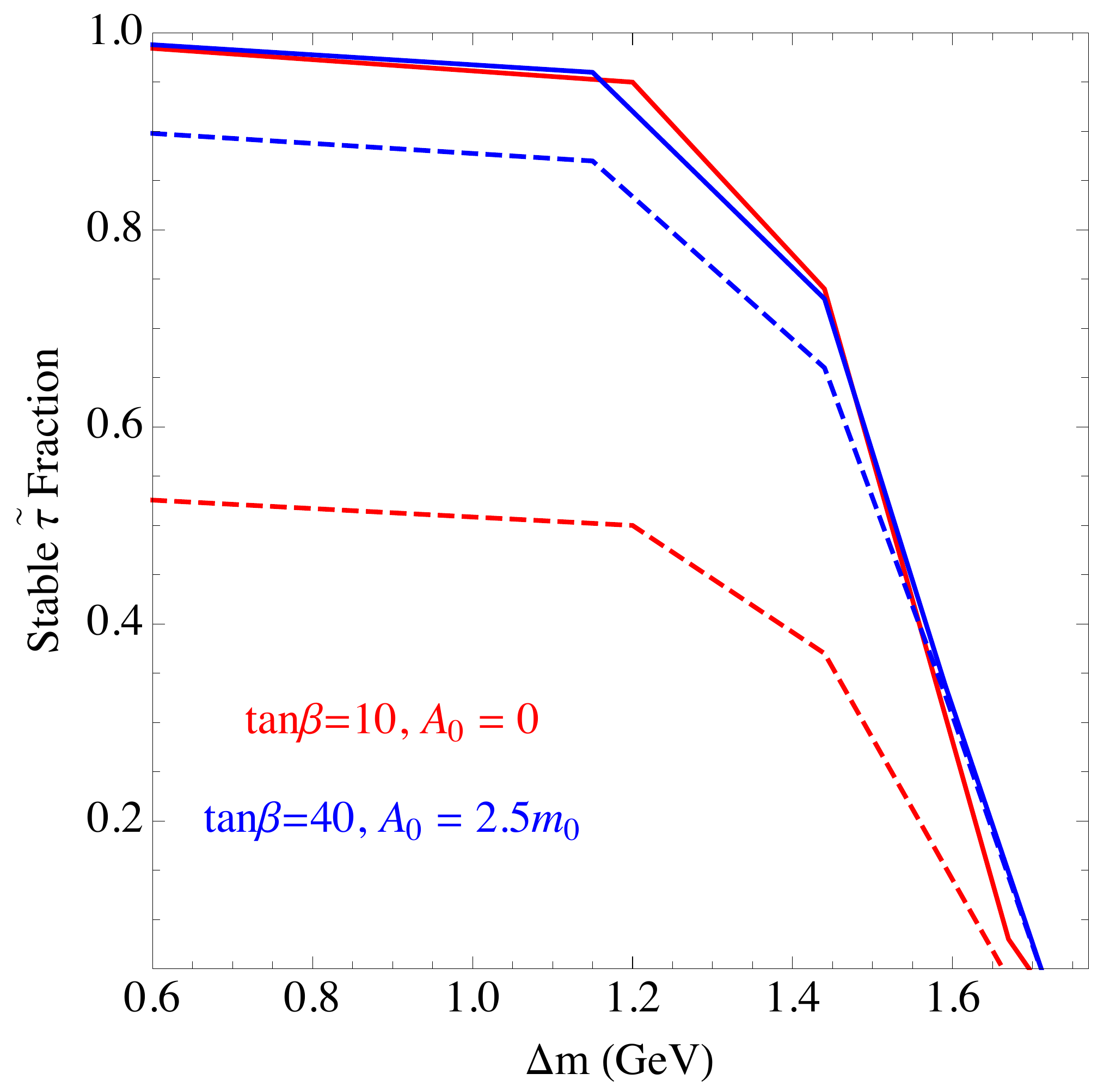,height=3.2in}
\epsfig{file=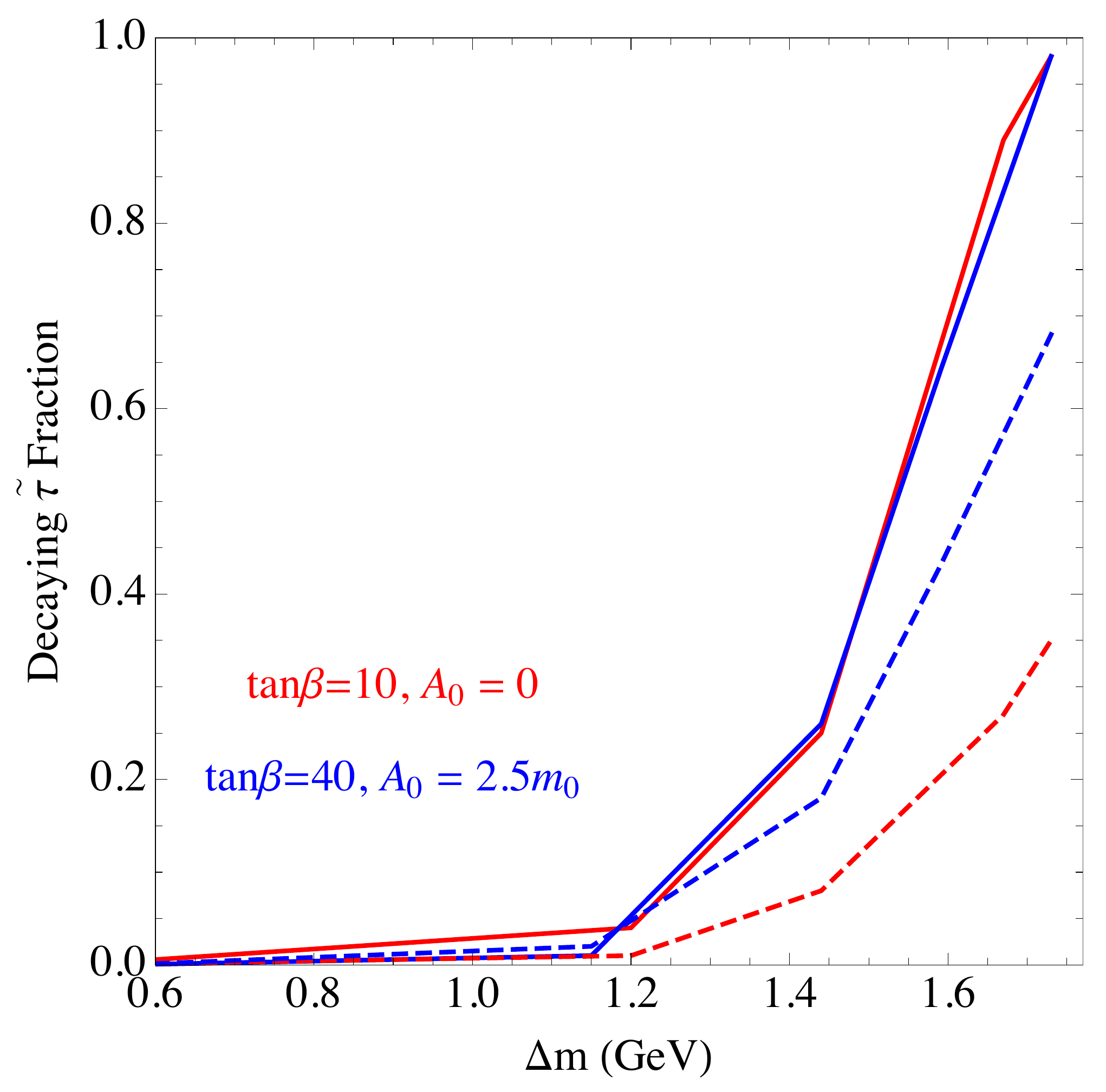,height=3.2in}
\end{minipage}
\caption{
{\it The dependence of the fraction of `stable' staus, i.e., those exiting CMS without decaying (left),
and the fraction of staus decaying within the ATLAS detector (right) on the $\tilde \tau - \chi$ mass difference.  The solid lines correspond to direct stau-pair production only whereas the dashed are for both direct and indirect stau production (i.e., all SUSY processes). The value of $m_{1/2} $ is fixed to 800 GeV.
}}
\label{fig:stable-decaying}
\end{figure}

The fraction of staus in the final state depends on the cascade decay branching ratios of the heavier sparticles,
which depend in turn on the CMSSM parameters, as seen in Fig.~\ref{fig:stable-decaying}.
For $\tan \beta = 40$ and $A_0 = 2.5 \,m_0$, for example, 
both the second lightest neutralino, $\chi_2^0$, and the lighter chargino, $\chi_1^\pm$, decay almost entirely into final states containing a stau,
whereas for $\tan \beta = 10$ and $A_0 = 0$ they decay into staus in only about 20\% and 64\%
of the cases, respectively.   However, even when the stau lifetime is long enough that most staus escape the detector, 
sensitivity to $E_T^{miss}$ is lost only if the decay chains of {\it both} the produced sparticles result in staus.
The dependence on $\Delta m$ of the fraction of staus stable enough to exit the CMS detector is shown in the left panel of 
Fig.~\ref{fig:stable-decaying}.  As expected, we find that for small $\Delta m$, most staus are stable.
The dashed lines (which correspond to the stau fraction from both direct and indirect production) asymptote to the total stau fraction in the final states as $\Delta m \to 0$.  As $\Delta m$ increases, the fraction of stable staus
decreases until it becomes zero at the tau mass threshold.
We find that when all production processes and decay chains are taken into account, 
the loss in efficiency for the $E_T^{miss}$-based search does not differ significantly for different values of $\tan \beta$ and $A_0$,
and the limit $m_{1/2} > 780$~GeV given in~\cite{ATLAS20} reduces to about 750 GeV.

We find that the most important constraint in the band where $\Delta m < m_\tau$ is that due to the
search for metastable charged particles, shown by the shaded blue regions in Fig.~\ref{fig:DeltaM}.
The darker shading is the constraint from direct stau pair production, and the lighter shading is
that obtained by including indirect stau production via the cascade decays of heavier sparticles.
The direct constraint yields a lower limit $m_{1/2} \gappeq 700$~GeV for $\Delta m \lappeq
1.4$~GeV, which is only weakly sensitive
to $\tan \beta$ but becomes stronger for $\Delta m \lappeq 1$~GeV when
$A_0 = 2.5 \, m_0$. The indirect constraint yields a lower limit that is somewhat
more sensitive to both $\tan \beta$ and $A_0$, yielding a lower limit $m_{1/2} \gappeq 800$
to 1100~GeV.

As can be seen from the right panel of Fig.~\ref{fig:stable-decaying},
the search for disappearing tracks becomes relevant for $m_\tau > \Delta m \gappeq 1.2$~GeV.
However, as already noted, it is weaker than the other constraints, yielding $m_{1/2} \gappeq 400$~GeV.
This is mainly because, even with a significant fraction of staus decaying before exiting the detector,
the signal efficiency for this search is of the order of 0.1-0.01\% after
implementing the cuts enumerated in section~\ref{sec:diss-track}.

In the case of $\tan \beta = 10$, for both $A_0 = 0$ and $2.5 \, m_0$ we see that the portions of the coannihilation strips
with $\Delta m \gappeq 3$~GeV are excluded by the ATLAS jets + $E_T^{miss}$ search, whereas the portions
with 3~GeV $\gappeq \Delta m > m_\tau$ are allowed by this search. For $\tan \beta = 40$
and $A_0 = 0 \, (2.5 \, m_0)$ the portion of the strip where 6~GeV (9~GeV)
$\gappeq \Delta m > m_\tau$ is allowed by this search. When $\Delta m < m_\tau$, we see that
none of the strips for $\tan \beta = 10, A_0 = 0$ or $\tan \beta = 40$ and $A_0 = 0$ or $2.5 \, m_0$ can be excluded, whereas
for $\tan \beta = 10$ and $A_0 = 2.5 \, m_0$ the portion of the strip with $\Delta m \lappeq 1.7$~GeV is excluded
by the search for massive charged particles.

In the ranges of $m_{1/2}$ exhibited in Fig.~\ref{fig:DeltaM}, {\tt FeynHiggs~2.10.0} generally yields values of $m_h$
below the value measured at the LHC. Taking into account the uncertainties in the theoretical calculation of $m_h$,
points yielding a nominal value $\sim 122.5$~GeV should probably not be regarded as excluded. 
Even taking this uncertainty into account, only the case $\tan \beta = 40, A_0 = 2.5 \, m_0$ is consistent
with the LHC measurement of $m_h$~\footnote{
We have checked that the value of $m_h$ calculated using
 {\tt SoftSUSY~3.3.7} is $\sim 1.5$~GeV smaller than the value given by {\tt FeynHiggs~2.10.0}
for the parameter ranges presented in Fig.~\ref{fig:DeltaM}. This difference is within the latter code's uncertainties.
}.

\section{The Potential Reaches of Future LHC Searches}

\begin{figure}[h!]
\hspace*{1.0in}
\begin{minipage}{8in}
\epsfig{file=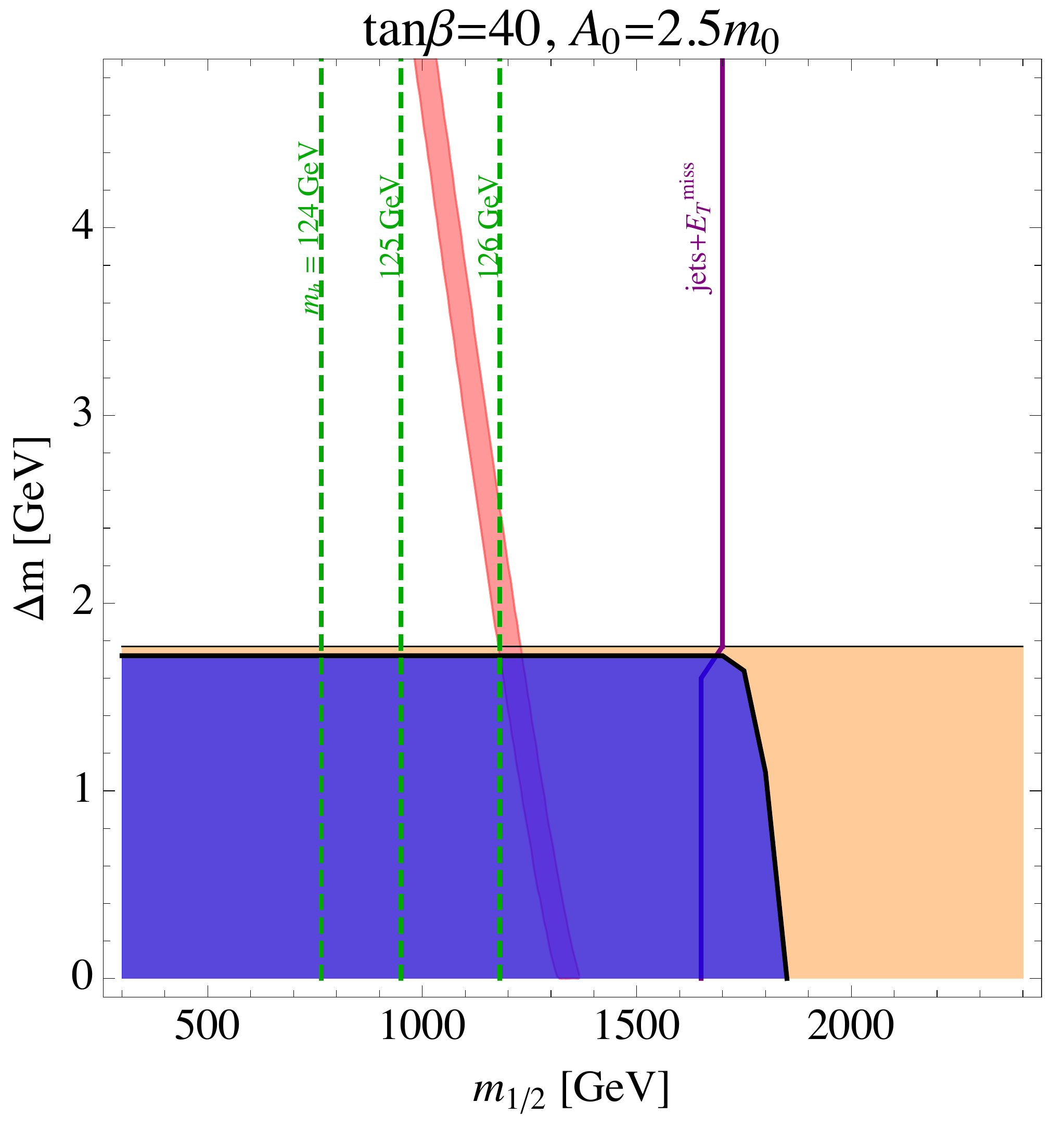,height=5in}
\end{minipage}
\caption{
{\it Projected limits from the 14 TeV LHC Run 2 with 300/fb integrated luminosity. 
The sensitivity of the jets + $E_T^{miss}$ search is sufficient to explore the rest of the coannihilation strip,
and the search for metastable charged tracks from direct stau-pair production is also strong 
enough to explore independently the portion of the coannihilation strip where $\Delta m \lappeq 1.6$~GeV.}}
\label{fig:lim14}
\end{figure}

In order to extrapolate the potential reach of each of the three distinct categories of searches 
used in the analysis of this paper, we make a simple but, we believe, reasonable assumption, 
namely that the expected cross-section sensitivities of the respective searches will remain the same when going 
from a centre-of-mass energy of 8 TeV to 13 or 14 TeV. Based on this assumption, 
we use {\sc Pythia~8}~\cite{pythia8, py8susy} to recalculate the cross-sections in this higher centre-of-mass regime, 
and extrapolate the mass limit by requiring that, at the new mass limit, 
the cross-section multiplied by the respective integrated luminosity is the same as for the Run 1 LHC data. 
The results of applying this hypothesis are shown in Fig.~\ref{fig:lim14}.

Fig.~\ref{fig:lim14} compares the prospective limits with the tip of the stau coannihilation strip
for $\tan \beta = 40$ and $A_0 = 2.5 \,m_0$, which was shown in Fig.~\ref{fig:DeltaM} to be
the most difficult to exclude. We do not display the projected sensitivity of
the `disappearing track' search, which we do not expect to be competitive because
of its low efficiency, as discussed earlier.
We see that the projected sensitivity of the jets + $E_T^{miss}$ search is $\sim 1700$~GeV
for $\Delta m > m_\tau$, decreasing to $\sim 1650$~GeV for $\Delta m < 1.6$~GeV, which is
sufficient to explore all the coannihilation strip.
The most sensitive search channel when $\Delta m \lappeq 1.6$~GeV is expected to be that
for massive charged particles, which would also be strong enough to explore independently this portion of
the coannihilation strip: its sensitivity should reach $\sim 1850$~GeV for very small $\Delta m$.
The combination of these searches would clearly explore thoroughly the CMSSM
coannihilation strip in this and, {\it a fortiori}, the other cases we study.
Indeed, we estimate the tip of the strip at $m_{1/2} \sim 1400$ GeV
can be explored with 75/fb data at 14 TeV. Points within the stau coannihilation
strip could be detected in two ways, via their $E_T^{miss}$ and massive stable particle
signatures. Conversely, the absence of signals in both these channels would exclude robustly the
stau coannihilation strip.

\section{Conclusions}

We have analyzed in this paper the impacts
on the CMSSM parameter space in the neighbourhood of the tip of
the stau coannihilation strip of various LHC searches, 
including the ATLAS jets + $E_T^{miss}$ search, the CMS search for
massive charged particles, and the ATLAS search for disappearing tracks,
which are sensitive in different regions of $\Delta m$ and $m_{1/2}$.
We have found that the jets + $E_T^{miss}$ search has important sensitivity
when $\Delta m < m_\tau$, though the strongest constraint for small $\Delta m$
is generally that provided by the search for massive charged particles.
The search for disappearing tracks has impact only for $\Delta m \gappeq 1.6$~GeV,
where it is considerably less sensitive than the jets + $E_T^{miss}$ search.

We have studied four CMSSM cases, with the following conclusions.
For $\tan \beta = 10$ and $A_0 = 0$, the portion of the coannihilation
strip with $\Delta m \gappeq 3$~GeV is excluded by the jets + $E_T^{miss}$
search, but the portion with $\Delta m \lappeq 3$~GeV cannot yet be excluded.
For $\tan \beta = 10$ and $A_0 = 2.5 \, m_0$, the portion of the coannihilation
strip with $\Delta m \gappeq 3$~GeV is again excluded by the jets + $E_T^{miss}$
search, and the portion with $\Delta m \lappeq 1.7$~GeV is excluded
by the search for massive charged particles, but there is no exclusion for the portion with
1.7~GeV $\lappeq \Delta m \lappeq 3$~GeV.
For $\tan \beta = 40$ and $A_0 = 0$, only the portion of the coannihilation
strip with $\Delta m \gappeq 6$~GeV is excluded, again by the jets + $E_T^{miss}$
search, and there is no exclusion for $\Delta m < m_\tau$.
Finally, for $\tan \beta = 40$ and $A_0 = 2.5 \, m_0$, only the portion of the coannihilation
strip with $\Delta m \gappeq 9$~GeV is excluded, again by the jets + $E_T^{miss}$
search.

We have also projected the likely sensitivities of the LHC searches in
Run 2 of the LHC at energies approaching 14~TeV and with up to
300/fb of integrated luminosity. We find that a combination of the
jets + $E_T^{miss}$ and massive charged particle searches should be
able to explore robustly the entire CMSSM coannihilation strip for all the
cases we have studied. The end of the CMSSM coannihilation strip
is indeed nigh, one way or another.

\section*{Acknowledgements}

This work was supported partly by the London
Centre for Terauniverse Studies (LCTS), using funding from the European
Research Council via the Advanced Investigator Grant 267352.
The work of J.E. was also supported in part by the UK STFC
via the research grant ST/J002798/1. N.D. would like to thank the Alexander von Humboldt Foundation for support.

\end{document}